\documentstyle[twocolumn,floats,aps,epsfig,prc]{revtex}
\include{graphics.sty}

\makeatletter

\providecommand{\LyX}{L\kern-.1667em\lower.25em\hbox{Y}\kern-.125emX\@}

\makeatother

\begin{document}

\draft

\title{The Radiation Tail in \protect\( (e,e'p)\protect \) Reactions
  and Corrections to Experimental Data}

\author{J. A. Templon}

\address{University of Georgia, Athens, GA 30621}

\author{C. E. Vellidis}

\address{University of Athens, Greece}

\author{R. E. J. Florizone}

\address{Massachusetts Institute of Technology, Cambridge, MA 02139}

\author{A. J. Sarty}

\address{Florida State University, Tallahassee, FL 32306}

\date{\today{}}

\maketitle 
We present a direct calculation of the cross section for
the reaction \( ^{3} \)He\( (e,e'p) \) including the radiation tail
originating from bremsstrahlung processes. This calculation is
compared to measured cross sections. The calculation is carried out
from within a Monte Carlo simulation program so that
acceptance-averaging effects, along with a subset of possible energy
losses, are taken into account.  Excellent agreement is obtained
between our calculation and measured data, after a correction factor
for higher-order bremsstrahlung is devised and applied to the tail.
Industry-standard radiative corrections fail miserably for these data,
and we use the results of our calculation to dissect the failure.
Implications for design and analysis of experiments in the
Jefferson-Lab energy domain are discussed.

\pacs{13.40.Ks,24.10.Lx,25.30.Fj}

\section{Introduction}

The application of what are commonly called \emph{radiative
  corrections} is an integral part of doing nuclear physics with beams
of electrons. In an electron-scattering experiment, the probe is
considered to be a virtual photon. This photon is exchanged between a
beam electron and a target nucleus, thereby transferring energy and
momentum to the target from the the electron, which is thereby
scattered. Unfortunately, these electrons also copiously emit real
photons which are not normally observed in experiments. Thus, either
the theoretical calculations with which data are compared must include
these processes (and they normally do not), or the data must somehow
be corrected for these effects so that they can be compared to
calculations which are based on single virtual-photon exchange.
The standard choice is to ``radiatively unfold'' the experimental
data, which generates a ``corrected spectrum'' that can be compared
to theoretical calculations.

This paper reports on a study of how these real-photon processes
affect measurements of \( (e,e'p) \) reactions on atomic nuclei.
Our calculation takes the second approach, which is to radiatively
correct \emph{a theoretical calculation} so that it can be
directly compared to uncorrected data.
We compare a direct computation of a cross section, including the effects
of photon emission, to a specific measurement of an \( (e,e'p) \)
cross section\cite{flor99,flor99a}. The results of applying the
standard ``radiative unfolding'' procedure mentioned above to these data
are also presented and discussed. The comparisons to this particular data
set are unique in two ways:

\begin{enumerate}
\item the data appear to be well described in the plane-wave impulse
  approximation (PWIA), and
\item over most of the kinematic range of the measurement, the
  real-photon ``radiative processes'' are responsible for nearly the
  entire cross section.
\end{enumerate}
The excellent PWIA reproduction of the cross section chosen for this
study allowed us to use the PWIA in carrying out the complex
calculations including radiative effects, enormously simplifying the
task. The dominance of radiative strength enables us to make a true
test of the real-photon emission model without worrying about
accurately removing physical backgrounds.

This study is timely for several reasons. First, existing procedures
for radiative corrections to data have been developed for experiments
at relatively low (\( <500 \) MeV) electron beam energy. Refinements
or overhauls of the procedure may be necessary to apply corrections
for \( (e,e'p) \) experiments with higher-energy beams. The experiment
studied here was carried out with a beam energy of 855 MeV, which
bridges the gap between the energy domain studied by the labs active
in the last decade (0.2--0.9 GeV) and the Jefferson Lab energy domain
(0.8--6.0 GeV). Furthermore, a new class of experiments at Jefferson
Lab begins to study \( (e,e'p) \) reactions in a kinematic domain
where the cross sections are expected to be small and broadly peaked;
radiative strength can easily swamp the ``true'' cross section in
these cases. The design and analysis of these experiments should make
careful studies of the radiative contributions to measured cross
sections. Indeed, such an analysis was the genesis for the current
work.

Finally, it became clear to us during the course of the project
described here that the standard radiative-unfolding procedure used
for the last decade is of an \emph{ad hoc} nature; it is not based on
rigorous theoretical arguments.  We could only find one article
published in a refereed journal \cite{deca91} which specifically
addressed radiative corrections for \( (e,e'p) \) reactions.  This
publication is either unknown to most experimentalists, or has been
ignored for some reason, as a literature search uncovered only one
reference \cite{maki94} to this article, which argued that the
corrections proposed in \cite{deca91} were impractical since they make
different assumptions about hadronic portions of the corrections,
rendering data so corrected inconsistent with the world proton
form-factor data.

All remaining works we could find addressing radiative corrections
to experimental data
were Ph.D.  theses. Essentially all these works quote the Ph.D. thesis
of Quint \cite{quin88} as the primary reference. This thesis in turn
quotes lecture notes of Penner \cite{penn77} from a summer-school
proceedings as a primary source, where radiative corrections for
inclusive \( (e,e') \) reactions are discussed. These notes clearly
state that the correction should be viewed as approximate; for
example, they recommend an empirical adjustment of the calculated tail
to give the best fit to the data, in cases where the radiation tail
dominates the cross section.  Aside from this problem and possible
problems in adapting a formalism for \( (e,e') \) to correct \(
(e,e'p) \) experiments, we have uncovered several questionable
assumptions in this standard procedure, which we address in this
article.

By contrast, we base our calculations on a published \cite{bori71}
first-order QED calculation for the radiation-tail cross section. Our
work extends their result to kinematically-complete reactions and to
higher-order bremsstrahlung radiation.

We are unaware of any previously-published similar study. We hope this
paper will give some indication of how urgently new theoretical work
is needed, and in what directions that work should proceed. Since the
topic of radiative corrections is often viewed as an arcane subject
which is best avoided, we present in the following sections a review
of the relevant electromagnetic processes, a review of \( (e,e'p) \)
phenomenology, and an explanation on how radiative effects distort \(
(e,e'p) \) reaction data.

\section{Review of Radiative Processes in Electron Scattering}

This section presents a review of the most important processes via
which electrons emit real photons during interactions with nuclei. We
emphasize here that this section is a review, meant to place these
processes in the context of the reaction we study. Much of the
conceptual work here, and formal work on bremsstrahlung presented in
sec.~\ref{sec:comput}, can be found in classic articles
\cite{schw49,mots69,tsai74}.

\vspace{0.3cm}
{
\begin{figure}
  {\par\centering
    \resizebox*{0.3\textwidth}{!}{\includegraphics{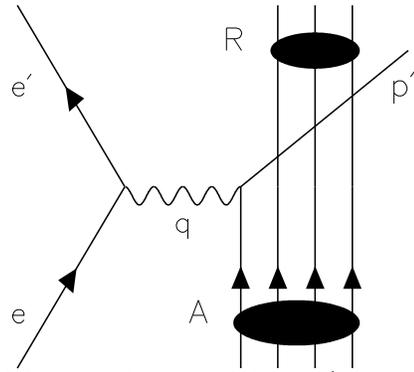}} \par}

\caption{Schematic diagram of the \protect\( (e,e'p)\protect \) reaction
  (in this case off an \protect\( A=4\protect \) nucleus) to lowest
  electromagnetic order.  An incident electron scatters from a target
  nucleus by exchange of a virtual photon, and a target proton is
  knocked out in the process. Symbols next to the various lines show
  the names given the four-momenta for each particle.}

{\par\centering \label{fig:feyn1}\par}
\end{figure}
\par}
\vspace{0.3cm}

\begin{figure}
  {\par\centering
    \resizebox*{0.2\textwidth}{!}{\includegraphics{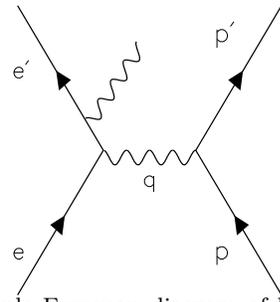}} \par}

\caption{Example Feynman diagram of bremsstrahlung in 
  \protect\( e\protect \)-\protect\( p\protect \) scattering. A real
  photon is emitted from the outgoing electron.  There are three other
  such diagrams, one each for the two proton legs and one for the
  incident electron leg.}

{\par\centering \label{fig:feyn2}\par}
\end{figure}

Fig.\ \ref{fig:feyn1} is a schematic diagram of the \( (e,e'p) \)
process to leading order in the electromagnetic coupling constant \(
\alpha \). It corresponds to the mental picture usually employed by an
experimentalist designing or analyzing an experiment, since it probes
the ``signal'' the experimenter usually wants to measure. It also
corresponds to the usual plane-wave impulse approximation (PWIA) for
\( (e,e'p) \) reactions. For the purpose of the study presented here,
we have chosen a measurement on \( ^{3} \)He which was performed in
kinematics specifically chosen to optimize the accuracy of the PWIA.
However, even in the limit that the PWIA holds for the hadronic
portion of the process, the neglect of real-photon emission limits the
accuracy of PWIA cross sections to at best 20\% for electron energies
above a few hundred MeV. An example diagram for this
\emph{bremsstrahlung} process is shown in Fig.~\ref{fig:feyn2}. Here
the spectator nucleons have been omitted from the picture for
simplicity.

\begin{figure}
  {\par\centering
    \resizebox*{0.2\textwidth}{!}{\includegraphics{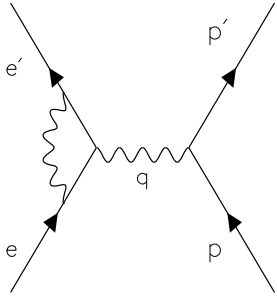}} \par}

\caption{Example diagram for the virtual photon correction.}

{\par\centering \label{fig:feyn3}\par}
\end{figure}

The process in Fig.~\ref{fig:feyn2} is called \emph{internal
  bremsstrahlung} since it occurs during the \( (e,e'p) \) reaction. A
similar process (termed \emph{external bremsstrahlung)} takes place in
the Coulomb fields of other atoms in the target. A related process is
not particularly relevant for understanding the effect of radiative
processes on the \( (e,e'p) \) spectra, but must be included in any
consistent calculation. This is the process in which two virtual
photons are emitted, and an example diagram is shown in
Fig.~\ref{fig:feyn3}.  Such diagrams are generally termed ``virtual
photon corrections.''

For fixed values of the four-momenta \( e \) and \( e' \), we see that
the value of the four-momentum transfer \( q \) is changed in the
diagram of Fig.~\ref{fig:feyn2} with respect
to the leading-order process in Fig.~\ref{fig:feyn1}. This in general
leads to a change in the magnitude of the associated cross section
(as does the vertex renormalization in Fig.~\ref{fig:feyn3}).
The extra emitted particle in Fig.~\ref{fig:feyn2} leads to a change
in the asymptotic kinematics of the reaction as well. This creates an
ambiguity; for a given measured event, it is impossible to tell
whether the observed kinematics correspond to those of the reaction
vertex, or to a different reaction-vertex situation accompanied by
real photon emission.

We begin our discussion of this problem by summarizing the
phenomenology of the \( (e,e'p) \) reaction in the next section. The
section afterwards will give a quantitative description of the
kinematic distortion due to the contributions of the diagram in
Fig.~\ref{fig:feyn2}. We present a rather complete summary of the
kinematics, since discrepant conventions exist in the literature. In
our discussions below, unless otherwise stated we use the following
kinematic conventions:

\begin{itemize}
\item the four-vectors are denoted by the standard symbol for the
  associated particle, \emph{e.g.} \( A \) for the target
  four-momentum and \( p' \) for that of the knocked-out proton.
\item \( E_{g} \) and \( {\bf p}_{g} \) refer to the relativistic
  energy and three-momentum of particle \( g \), thus the four
  momentum for \( g \) is \( g=(E_{g},{\bf p}_{g}) \).  The magnitude
  of the three momentum is \( p_{g}=\left| {\bf p}_{g}\right| \).
\end{itemize}

\section{Review of \textit{}\lowercase{$(e,e'p)$} \textit{}Phenomenology}

\label{sec:eephen}One of the main reasons why the \( (e,e'p) \) reaction has
been so useful in nuclear physics is that it probes the properties of
individual nuclear protons in a fairly direct manner. The measured
particle momenta can be used to determine the energy and momentum that
the struck proton had before the interaction. The only major
assumption involved is that the interactions of the recoiling \( (A-1)
\) system and knocked-out proton are neglected (PWIA).  While the PWIA
is not sufficiently accurate for a quantitative analysis of \( (e,e'p)
\) experiments, many experiments have shown \cite{lapi93} that the
essential features are reasonably preserved and that a straightforward
analysis is possible.

The probability of finding the ``struck'' proton of
Fig.~\ref{fig:feyn1} (to which the four-momentum \( q \) is
transferred) is a function of two parameters: an energy (which we
refer to as \( \epsilon \)) and a momentum (to which we refer as \(
\bbox {\rho } \)). Various equivalent conventions for \( \epsilon \)
exist; we use it to refer to the energy necessary to remove the proton
from the nucleus. \( \bbox {\rho } \) refers to the momentum of the
proton relative to the nuclear rest frame.

In our convention, \( \epsilon \) consists of two parts: \( \epsilon
=S_{p}+E_{x}^{R} \).  \( S_{p} \) is the proton separation energy for
the nucleus being bombarded, and \( E_{x}^{R} \) is the excitation
energy of the residual system ``\( R \)'' of \( A-1 \) nucleons.

Assuming that the PWIA holds, \emph{\( \epsilon \)} and \( \bbox {\rho
  } \) can be computed from the kinematic variables measured in \(
(e,e'p) \) experiments.  We begin by constructing a four-vector
relation for the process (using the notation of Fig.~\ref{fig:feyn1}):
\begin{equation}
\label{eq:fvpwia}
e+A=e'+R+p'.
\end{equation}
Assuming the reaction is carried out with a known beam energy, a
fixed, pure target, and that the four-momenta of the scattered
electron and knocked-out proton are measured in detectors, the
kinematics are uniquely determined.
\begin{equation}
\label{eq:fv_r}
R=(E_{R},{\bf p}_{R})=(e-e')+A-p'.
\end{equation}
The invariant mass of the \( (A-1) \) system, \( \sqrt{R^{2}} \),
yields \( m_{R} \); an experimental \emph{missing energy}\footnote{
  Some authors use \( E_{m} \) to denote the \emph{unmeasured}
  ``missing'' energy in the reaction, which thus includes the kinetic
  energy of the recoiling undetected system. This terminology is
  historically correct, since in early experiments with low-energy
  beams on heavy targets, the recoil kinetic energy was negligible. \(
  E_{m} \) became synonymous with the binding energy. Later,
  approximate corrections were used to remove the recoil energy. The
  use of relativistic invariants eliminates the need for
  approximations. Our value might be more properly termed the
  ``missing mass'' since all other forms of energy have been accounted
  for; nevertheless, we stick with the historical term.  } is computed
as
\begin{equation}
\label{eq:emdef}
E_{m}=m_{R}+m_{p}-m_{A}.
\end{equation}
When PWIA holds, \( \epsilon =E_{m} \). Similarly, an experimental
\emph{missing momentum} is defined as
\[
{\bf p}_{m}={\bf p}_{R}\, .\] When PWIA holds, the residual system is
a spectator and thus must have had the same momentum \( {\bf p}_{m} \)
before the interaction. Since the nucleus as a whole was initially at
rest, \( \bbox {\rho }=-{\bf p}_{m}=-{\bf p}_{R} \).

\section{The Kinematical Effects of Bremsstrahlung}

\label{sec:kineff}If we add a real photon as shown in Fig.~\ref{fig:feyn2}
to one of the external legs in Fig.~\ref{fig:feyn1}, we must account
for it in the four-momentum conservation relation. Here we keep using
\( E_{m} \) and \( p_{m} \) for the names of the measured quantities.
We indicate by use of the extra subscript $v$
(\emph{e.g.,}\ \({\bf p}_{m,v} \))
the corresponding quantity at the $q A$ (virtual-photon-nucleus)
reaction vertex
in the case that the actual reaction involved emission of a real
photon.

The four-momentum conservation relation becomes 
\begin{equation}
\label{eq:rwphot}
R=(e-e')+A-p'-\gamma ,
\end{equation}
where \( \gamma =(k,{\bf k}) \) refers to the real photon's
four-momentum.  The three-vector component of this equation yields
\begin{eqnarray}
{\bf p}_{R,v}   & = & {\bf q}-{\bf p}_{p'}-{\bf k}\mbox{, so that}
                                               \label{eq:phot3v} \\
{\bf p}_{m,v} & = & {\bf p}_{m}-{\bf k}\, .
\end{eqnarray}
Thus the deduced missing momentum is \( {\bf p}_{m}={\bf
  p}_{m,v}+{\bf k} \).  The effect of the real photon emission on \(
E_{m} \) is not obvious when using four-momentum algebra; instead, we
note that the zeroth component of (\ref{eq:rwphot}) leads to
\begin{equation}
\label{emwphot}
m_{R,v}+m_{p}-m_{A}=(E_{e}-E_{e'}-T_{p'}-T_{R,v})-k,
\end{equation}
where \( T \) refers to the particle's \emph{kinetic} energy and \( k
\) is the photon energy. The left-hand side of this relation is \(
E_{m,v} \), and the value in parentheses is, to a good approximation,
what one would \emph{deduce}
for \( E_{m} \) if one is ignorant of the photon emission. Thus \(
E_{m}\approx E_{m,v}+k \).  The relation is approximate since
one measures \( T_R \), not \( T_{R,v} \).  However, the
difference is in most cases quite small.

\begin{figure*}
  {\par\centering
    \resizebox*{0.6\textwidth}{!}{\includegraphics{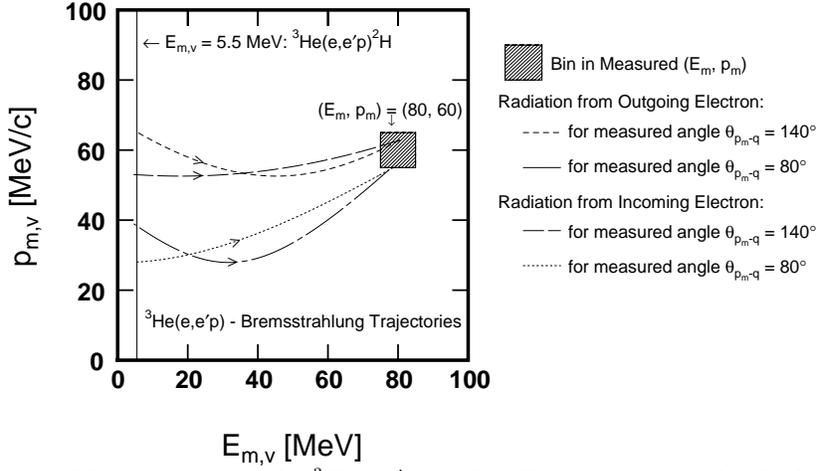}}
    \par}

\caption{\label{fig:adam1}
  Example bremsstrahlung trajectories for
  \protect\( ^{3}\protect \)He\protect\( (e,e'p)\protect \) in the
  \protect\( (E_{m,v},p_{m,v})\protect \) plane for the kinematic settings
  given in Table \ref{tab:kinset}. Each line represents a trajectory
  through which strength from lower values of \protect\( E_{m,v}\protect
  \) can feed into the selected bin of measured \protect\(
  (E_{m},p_{m})\protect \).  The four lines show paths
  for radiation occurring either on the incoming or outgoing electron
  leg, and also for two values of the deduced measured angle between
  \protect\( {\bf p}_{m}\protect \) and \protect\( {\bf q}\protect \),
  \protect\( \theta _{p_{m}-q}\protect \).}
\end{figure*}

The bremsstrahlung-photon emission thus causes cross-section strength
which would normally populate \( (E_{m,v},p_{m,v}) \) to instead be
redistributed over a range of values \( (E_{m},p_{m}) \).
Furthermore, the magnitude of this redistributed cross section
will be modified since the momentum transfer \( q \) will be changed.
The redistributed strength is the origin of the well-known ``radiation
tail'' of electron-scattering experiments. In general, the missing
energy is simply increased by an amount equal to the radiated-photon
energy.  The missing momentum is shifted in a kinematic-dependent
manner; the relative orientation of its vertex value \(
{\bf p}_{m,v} \) and that of the radiated photon \( {\bf
  k} \) plays an important role. Fig.~\ref{fig:adam1} illustrates how
measured strength in a particular region of \(
(E_{m},p_{m}) \) is fed, through the bremsstrahlung
process, by various regions of \( (E_{m,v},p_{m,v}) \).

A procedure (see Ref.~\cite{holt95} for a good discussion) has been
developed for ``radiatively correcting'' \( (e,e'p) \) spectra. The
procedure relies on the fact that in reactions for which \( E_{m} \)
is a minimum, there is no tail, only a reduction in the cross section
due to the absence of that strength which has been moved into
the tail at larger
values of \( E_{m} \). The procedure corrects cross section data by
beginning with this minimum-\( E_{m} \) bin, using the Schwinger
correction to correct for the amount which has been lost to the tail.
Then the tail itself is computed from this effectively ``unradiated''
cross section. The tail contribution from this bin is subtracted from
all bins at larger missing energies. This procedure can be iterated
bin-by-bin, moving from small \( E_{m} \) to large, to remove the
radiation tail.

Estimated uncertainties in the \emph{value} of the correction are
usually around 10--20\%, which is acceptable when the correction
itself is small. However, for experiments investigating the large-\(
E_{m} \) continuum cross section, the correction can become rather
large, or the radiation tail can even dominate the cross section. In
these cases, even a 10\% uncertainty in the corrections can lead to
essentially zero knowledge of the true ``unradiated'' cross section.
It is important to be able to reliably estimate the strength of the
radiation tail, so cases like this can be avoided during the planning
stage of an experiment.  In the following section, we describe the
procedure for computing the cross section, including the radiation
tail.

\section{Computation of \lowercase{$(e,e'p)$} Cross Sections 
  Including Radiative Processes}

\label{sec:comput}Our approach to direct computation of the \( (e,e'p) \)
cross-section spectra, including bremsstrahlung processes, is based on
the PWIA for the \( (e,e'p) \) reaction. The PWIA assumption is not a
necessary one; however, a computation involving a more complete theory
would be computationally much more intensive. The use of PWIA is
well-motivated in our case since it works well for \( ^{3} \)He\(
(e,e'p) \), apart from an overall scaling factor; this has been
observed in other experiments as well (see \emph{e.g.} \cite{jans87}).
The basic program is as follows: a Monte-Carlo simulation is performed
in which the four-momenta of the scattered electron and knocked-out
proton are sampled over their respective acceptances. The kinematics
for the \( (e,e'p) \) reaction vertex are then modified for
bremsstrahlung processes according to the corresponding probability
distributions. PWIA is finally used to compute the cross section, and
any relevant jacobian factors are applied.

The data generated can then be used to form cross-section spectra
using the same procedures applied when analyzing the experimental
data. The use of Monte-Carlo simulation is important for an additional
reason: the cross sections are often rapidly varying over the
experimental acceptances. A calculation of cross-section spectra using
only the kinematics corresponding to the centers of all the
acceptances does not usually reproduce the experimental spectra.
Indeed, when radiative corrections are applied to ``deradiate''
experimental data, often one of the biggest uncertainties stems from
the acceptance-averaging assumption made. We will discuss this point
in more detail later in the article. Our Monte-Carlo procedure allows
for acceptance-averaging identical to that of the experiment.

\subsection{PWIA Cross Sections}

The cross section for continuum \( (e,e'p) \) reactions, in the case
where the residual \( (A-1) \) system has a continuum of possible
invariant masses \( m_{R} \) is given by
\begin{equation}
\label{eq:sigpw}
\frac{d^{6}\sigma }{d\Omega _{e'}d\omega d\Omega _{p'}dE_{p'}}=
  p_{p'}E_{p'}\sigma _{ep}S(E_{m},p_{m}).
\end{equation}
\( \omega \) is the electron energy loss, given in the laboratory
system by \( \omega =E_{e}-E_{e'} \); equivalently it is the zeroth
component of the four-vector momentum transfer \( q=(\omega ,{\bf q})
\). \( \sigma _{ep} \) is the elementary cross section for scattering
of an electron from a moving nucleon. We used the ``cc1'' prescription
\cite{defo83a} for this cross section.  The differences between
``cc1'' and ``cc2'' at these kinematics is everywhere less than 3\%
\cite{flor99}. Our calculation uses the Simon parameterization
\cite{simo80} of the nucleon electromagnetic form factors in the
computation of \( \sigma _{ep} \). \( S(E_{m},p_{m}) \) is the proton
spectral function, which gives the probability of finding a proton in
the nucleus with momentum \( p_{m} \) and removal energy \( E_{m} \).

The simulation samples particle momenta instead of energies, so the
cross section must be adjusted before direct use by the program. For
the continuum case, we have to apply the transformations \( \omega
\Rightarrow p_{e'} \) and \( E_{p'}\Rightarrow p_{p'} \).  The
resulting cross section used in the simulation is related to that of
Eq.~(\ref{eq:sigpw}) by
\begin{equation}
\label{eq:sixf-xf}
\frac{d^{6}\sigma }{d\Omega _{e'}dp_{e'}d\Omega _{p'}dp_{p'}}=
\frac{p_{p'}}{E_{p}'}
\frac{d^{6}\sigma }{d\Omega _{e'}d\omega d\Omega _{p'}dE_{p'}}.
\end{equation}

In the case of reactions leaving the residual system in a discrete
state, the cross section is given by
\begin{equation}
\label{eq:pw2b}
\frac{d^{5}\sigma }{d\Omega _{e'}d\Omega _{p'}dp_{e'}}=
\frac{p_{p'}E_{p'}\sigma ^{\textrm{cc}1}_{ep}n(p_{m})}{R}.
\end{equation}
For these reactions \( E_{m} \) has a definite value \( E_{\alpha }
\), so \( S(E_{m},p_{m}) \) is replaced by the momentum distribution
\( n(p_{m}) \), where
\begin{equation}
\label{eq:specdiscrete}
S(E_{m},p_{m})=n(p_{m})\delta (E_{m}-E_{\alpha }).
\end{equation}
\( R \) is the ``recoil factor'' (really a jacobian factor
transforming \( E_{p'}\Rightarrow E_{m} \)), and is given by
\begin{equation}
\label{recofac}
R=1-\frac{E_{p'}}{E_{R}}
\frac{{\bf p}_{R}\cdot {\bf p}_{p'}}{\left| {\bf p}_{p'}\right| ^{2}}.
\end{equation}
 
Here we have not included the extra subscript $v$ on the kinematic
quantities, but it should be understood when evaluating the
radiative cross sections later in this section, that the
hadronic cross section terms must be evaluated at the hadronic-vertex
kinematics.  Subscripts have been added in that section as a
reminder.

So far, we have only studied light nuclei with only one possible
discrete transition (the \( A-1 \) ground state), so our complete
simulation consists of a sum of one discrete simulation and one
continuum simulation. These two simulations are each themselves
composed of two simulations to handle the different pieces of the
radiation tail. The procedure is straightforward to extend to more
complicated situations including additional bound-state channels.

\subsection{External Bremsstrahlung}

External bremsstrahlung is relatively simple to include. The
simulation program contains a model for the actual reaction target.
For each event, an interaction vertex is chosen randomly within the
intersection of the beam and target volumes.  After this choice, the
amount of target material traversed by the electron before and after
the reaction is computed. The electron is transported through each
material (\emph{e.g.,}\ target cell walls or target gas). After each
traversal, a new electron energy is generated directly from sampling
Tsai's distribution \cite{tsai74} for bremsstrahlung from a thin
radiator:
\begin{eqnarray}
I_{\textrm{ext}}(E_{0},k ,t) & = & 
   \frac{bt}{\Gamma (1+bt)}
   \left( \frac{k }{E_{0}}\right) ^{bt}
   \times \nonumber \label{eqn:tsai1} \\
 &  & \frac{1}{k }
   \left( 1-\frac{k }{E_{0}}+\frac{3}{4}
   \left( \frac{k }{E_{0}}\right) ^{2}\right) .
\end{eqnarray}
Here \( k \) is the radiated photon energy (or energy lost by
the electron), \( E_{0} \) is the energy of the electron upon entering
the radiator material, \( t \) is the thickness of the radiator
material in radiation-length units, \( b \) is Tsai's bremsstrahlung
parameter (see Eq. (4.3) of \cite{tsai74}), and \( \Gamma \) is the
usual gamma function.

\subsection{Internal Bremsstrahlung}

Internal bremsstrahlung is included using the cross sections for
first-order photon emission derived by Borie and Drechsel
\cite{bori71}. Their derivation made use of the \emph{peaking
  approximation,} which assumes that bremsstrahlung photons are only
emitted along either the incident beam direction, or the direction of
the scattered electron's momentum. A critical review of the peaking
approximation can be found in \cite{mots69}. We use their results to
estimate the validity of the peaking approximation for our kinematics,
and find that it should be accurate to better than 1\%. We note here
that it is difficult to make blanket statements about the peaking
approximation, except that it becomes increasingly worse for larger
bremsstrahlung-photon energies.

The Borie-Drechsel cross section was also derived specifically for \(
(e,e'p) \) reactions to the \( (A-1) \) continuum. Part of the present
work is an extension of that formalism to processes in which the \(
(A-1) \) system is in a discrete state. We also present a derivation
of a correction factor which accounts for higher-order bremsstrahlung
processes.

For the continuum case, there is complete kinematical freedom for all
particles, as long as the invariant mass of the \( (A-1) \) system is
large enough to be above the particle-emission threshold of the \(
(A-1) \) nucleus. The simulation then samples all kinematic variables
(the scattered-electron three-momentum, the ejected proton
three-momentum, and the emitted photon momentum). In this case, the
relevant cross section is given by Ref.~\cite{bori71}, but we repeat
it here with different notation and in a form consistent with our
results for the discrete case.
Unless otherwise specified, all kinematic quantities refer to the
asymptotic situation, \emph{i.e.,}\ what would be assigned if one
was not aware that a real photon had been emitted.
These cross sections have a two-term
structure which arises from the peaking approximation. The first term
below corresponds to ``preradiation'' (photon emission along the beam
direction) and the second term corresponds to ``postradiation''.
\begin{eqnarray}
\frac{d^{7}\sigma }{dkd\omega dp_{p'}d\Omega _{e'}d\Omega _{p'}} & = & 
   f_{\mathrm{mp}}\Bigl \{\Bigl (\frac{dr}{dk}\Bigr )_{e}\, \, 
   \frac{d^{6}(e-\gamma ,e')}{d\omega_v dp_{p'}d\Omega _{e',v}d\Omega _{p'}}
   \nonumber \label{eq:bd-cont} \\
 &  & +\Bigl (\frac{dr}{dk}\Bigr )_{e'}\, \, 
   \frac{d^{6}(e,e'+\gamma )}{d\omega_v dp_{p'}d\Omega _{e',v}d\Omega _{p'}}
   \Bigr \}.\label{eq:bd-cont} 
\end{eqnarray}
The factor \( f_{\mathrm{mp}} \) corresponds to our ``multi-photon''
correction factor which we will discuss later; \( f_{\mathrm{mp}}=1 \)
corresponds to the result published in \cite{bori71}. In this section,
we can use \( k \) for both the energy and the momentum of \( \gamma
\) since it is a real photon.  The \( dr/dk \) terms are essentially
jacobian factors for the photon emission.  They will be given below.
Finally, the two cross sections on the right-hand side of
Eq.~(\ref{eq:bd-cont}) are the usual ``unradiated'' cross sections,
and must be evaluated at the \emph{vertex} values.
This is why
for example the first cross section is a function of $e - \gamma$
(the beam four-momentum adjusted for photon emission before the
interaction) rather than of $e$.

For the discrete case, the kinematics are over-determined. Since both
the scattered electron and ejected proton are ``detected'' in the
simulation, but the photon is not, we sample over the six-dimensional
\( ({\bf p}_{e'},{\bf p}_{p'}) \) space. For each point in this space,
photon energies can be chosen which belong to this coordinate and as
well result in the correct invariant mass of the \( (A-1) \) system.
\( k_{e} \) is the real-photon energy in the case that the photon is
emitted along the direction of the incident electron, and \( k_{e'} \)
is that for the case of photon emission along the scattered-electron
direction. These values are in general not the same (as opposed to the
continuum case of Eq.~(\ref{eq:bd-cont}), where the values of \( k \)
\emph{were} the same).
\begin{equation}
k_{e}=\frac{\Lambda^{2}-m_{R}^{2}+2{\bf p}_{e}\cdot 
   ({\bf p}_{p'}+{\bf p}_{e'}-{\bf p}_{e}/2)-({\bf p}_{p'}+{\bf p}_{e'})^{2}
   }{2[\Lambda+\widehat{{\bf p}}_{e}\cdot ({\bf p}_{p'}+
   {\bf p}_{e'}-{\bf p}_{e})]}
\end{equation}
\begin{equation}
k_{e'}=\frac{\Lambda^{2}-m_{R}^{2}-2{\bf p}_{e'}\cdot 
   ({\bf p}_{p'}+{\bf p}_{e'}/2-{\bf p}_{e})-({\bf p}_{p'}-{\bf p}_{e})^{2}
   }{2[\Lambda+\widehat{{\bf p}}_{e'}\cdot ({\bf p}_{p'}+
   {\bf p}_{e'}-{\bf p}_{e})]}
\end{equation}
\begin{equation}
\Lambda=m_{A}+\omega -E_{p'}
\end{equation}
\begin{equation}
\widehat{{\bf p}}_{e,e'}=\frac{{\bf p}_{e,e'}}{|{\bf p}_{e,e'}|}
\end{equation}
Again here $\omega$ refers to the \emph{observed}, asymptotic value,
not that at the vertex.  Similarly, $\Lambda$ includes the total
energy of both the recoiling hadronic system and the radiated photon
since $\omega$ is the asymptotic value.

The associated cross section is
\begin{eqnarray}
\frac{d^{6}\sigma }{d\omega dp_{p'}d\Omega _{e'}d\Omega _{p'}} & = & 
   f_{\mathrm{mp}}\Bigl \{\Bigl |\frac{dk}{dp_{p'}}\Bigr |_{e}\Bigl (
   \frac{dr}{dk}\Bigr )_{e}\frac{d^{5}\sigma (e-\gamma _{e},e')
   }{d\omega_v d\Omega _{e',v}d\Omega _{p'}}\nonumber \label{eq:bd-2b} \\
 &  & +\Bigl |\frac{dk}{dp_{p'}}\Bigr |_{e'}\Bigl (\frac{dr}{dk}\Bigr )_{e'}
   \frac{d^{5}\sigma (e,e'+\gamma _{e'})}{d\omega_v d\Omega _{e',v}
        d\Omega _{p'}
   }\Bigr \}.\label{eq:bd-2b} 
\end{eqnarray}
\( f_{\mathrm{mp}} \) has the same meaning as in the preceding
paragraph.  \( \gamma _{e,e'} \) are the four-momenta corresponding to
\( k_{e,e'} \).  The above result (setting \( f_{\mathrm{mp}}=1 \) for
the moment) was generated by substituting Eq.~(\ref{eq:sigpw}),
coupled with the discrete-final-state expression for the spectral
function (Eq.~(\ref{eq:specdiscrete})), into the Borie-Drechsel
formula for the radiation tail (Eq.~(\ref{eq:bd-cont})). The integral
over \( dE_{m,v} \) was formally carried out by converting it, with the
help of appropriate jacobian factors, to an integral over \( dk \).
The kinematical factors for photon emission, one each for pre- and
postradiation, are given by
\begin{eqnarray}
\Bigl (\frac{dr}{dk}\Bigr )_{e} & = & \frac{\alpha }{\pi k_{e}}\, \, 
   \frac{E_{e}^{2}+(E_{e}-k_{e})^{2}}{E_{e}^{2}}\ln \frac{2E_{e}}{m_{e}}\\
\Bigl (\frac{dr}{dk}\Bigr )_{e'} & = & \frac{\alpha }{\pi k_{e'}}\, \, 
   \frac{(E_{e'}+k_{e'})^{2}+E_{e'}^{2}}{(E_{e'}+k_{e'})^{2}}\ln 
   \frac{2E_{e'}}{m_{e}}
\end{eqnarray}
In the continuum case, \( k_{e} \) and \( k_{e'} \) are identical (the
sampled photon energy). The jacobian factors transforming the cross
section from differential in \( k_{e,e'} \) to differential in \(
p_{p'} \) are
\begin{equation}
\Bigl |\frac{dk}{dp_{p'}}\Bigr |_{e,e'} = 
   \frac{A_{e,e'}+C_{e,e'}p_{p'}/E_{p'}}{B_{e,e'}-C_{e,e'}}
\end{equation}
 where 
\begin{equation}
A_{e}=p_{p'}+\widehat{{\bf p}}_{p'}\cdot 
   [{\bf p}_{e'}-({\bf p}_{e}-{\bf k}_{e})]
\end{equation}
\begin{equation}
A_{e'}=p_{p'}+\widehat{{\bf p}}_{p'}\cdot 
   [({\bf p}_{e'}+{\bf k}_{e'})-{\bf p}_{e}]
\end{equation}
\begin{equation}
B_{e}=(E_{e}-k_{e})-\widehat{{\bf p}}_{e}\cdot 
   ({\bf p}_{e'}+{\bf p}_{p'})
\end{equation}
\begin{equation}
B_{e'}=-(E_{e'}+k_{e'})+\widehat{{\bf p}}_{e'}\cdot 
   ({\bf p}_{e}-{\bf p}_{p'})
\end{equation}
\begin{equation}
C_{e,e'}=\Lambda-k_{e,e'}
\end{equation}
\begin{equation}
\widehat{{\bf p}}_{p'}=\frac{{\bf p}_{p'}}{|{\bf p}_{p'}|}
\end{equation}

\subsection{Schwinger Correction}

\label{sec:schw-corr}The internal-bremsstrahlung cross section given above
becomes singular as the radiated-photon energy goes to zero. Hence it
can not be used to provide the complete radiated cross section. The
classic technique is to choose a cutoff energy \( \Delta E \) which is
comparable to the experimental energy resolution; a radiation tail is
generated with photon energies between \( \Delta E \) and the full
energy of the radiating electron. The remaining cross section in the
originating kinematic bin (\emph{i.e.,} the cross section for this
particular reaction where the total energy radiated away by real
photons is less than \( \Delta E \)) is calculated by computing the
cross section without the internal bremsstrahlung graphs, and then
reducing this cross section to account for that strength which was
moved into the radiation tail. This reduction factor is called the
\emph{Schwinger correction}. For brevity, in discussions below we will
refer to the strength remaining in the original kinematic bin as the
``unradiated strength''.

The formalism we use for the Schwinger correction is due to Penner
\cite{penn77}, and is written as
\begin{equation}
\label{eq:schw}
C_{\mathrm{S}chw}=e^{-\delta _{r}}(1-\delta _{v}),
\end{equation}
where \( \delta _{r} \) is the first-order correction for internal
bremsstrahlung.  Penner's formulation is based on that of Maximon
(the expression at the bottom of p.\ 199 of ref.~\cite{maxi69})
with the addition of kinematic recoil corrections proposed by Tsai
\cite{tsai61}.
Furthermore, the part of this correction corresponding to
real-photon emission (\( \delta_r \)) has been exponentiated.
Exponentiation of this first-order correction was
suggested by Schwinger \cite{schw49} as a means of accounting for
higher-order (multiple-photon) bremsstrahlung.  \( \delta _{v} \) is
the correction for virtual-photon loops at the reaction vertex. The
two \( \delta \) factors are given by
\begin{eqnarray}
\delta _{r} & = & \frac{\alpha }{\pi }
   \Bigl (\ln \frac{Q^{2}}{m_{e}^{2}}-1\Bigr )\ln 
   \Bigl [\frac{\kappa }{\zeta ^{2}}
   \frac{E_{e}E_{e'}}{(\Delta E)^{2}}\Bigr ]\label{eq:delta} \\
\delta _{v} & = & \frac{\alpha }{\pi }
   \Bigl [\frac{28}{9}-\frac{13}{6}\ln 
   \frac{Q^{2}}{m_{e}^{2}}+\frac{1}{2}\ln ^{2}
   \frac{E_{e}}{E_{e'}}+\frac{\pi ^{2}}{6}\nonumber \\
 &  & -L_{2}\Bigl (\cos ^{2}\frac{\theta _{e'}}{2}\Bigr )\Bigr ]
\end{eqnarray}
 with the ``recoil factors'' given by 
\begin{eqnarray}
\zeta  & = & 1+\frac{E_{e}}{M_{A}}(1-\cos \theta _{e'})\text 
   {\, and}\label{eq:kapzet} \\
\kappa  & = & 1+\frac{\omega }{M_{A}}(1-\cos \theta _{e'})\, .
\end{eqnarray}

\( L_{2}(x) \) is the Spence function defined by
\[
L_{2}(x)=-\int _{0}^{x}\frac{\ln (1-y)}{y}dy\, .\]
Finally, \( Q^{2} \) denotes the standard square of the four-momentum
transfer, \( Q^{2}=-\left( e-e'\right) ^{2} \).

This version of the Schwinger correction does not account for possible
real-photon emission by the hadrons involved in the reaction.
Makins \cite{maki94} has noted that this process may begin to become
important momentum transfers $Q^2 > 1$ GeV/\emph{c}.
Penner's correction also omits all hadron self-energy and
vertex-renormalization diagrams.
The assumption implicit in this approach is that such diagrams become
part of what one calls the ``electromagnetic form factor''
of the struck hadron.

\section{Simulation Framework}

The models for \( (e,e'p) \) cross sections in PWIA, for external
bremsstrahlung, and for internal bremsstrahlung were implemented in
the simulation code AEEXB \cite{uga-97-2,vell97}. This code also
includes facilities enabling a fairly complete simulation of
experimental factors such as target geometry, beam energy dispersion,
ionization energy losses, and experimental acceptances. All these
facilities were used in order to make the comparison as realistic as
possible.  For practical reasons, certain classes of ionization energy
losses were not included. Since our electron energies are above the
critical energy (for which radiative energy loss processes become more
important than those due to atomic ionization), the ionization losses
had a negligible effect on our results. Ion-optical magnetic transport
is also possible in AEEXB, using an interface to the standard
ion-optics program TURTLE \cite{vell97,care82}, but was neither
necessary nor used for the current project.

A complete cross-section simulation including the radiation tail
consists of several distinct pieces which must be combined at the end
to obtain the final result. The framework is sketched here; readers
wishing to see a more detailed explanation should refer to
\cite{uga-98-1}. The discussion below makes the simplifying assumption
that the final state space for the residual \( (A-1) \) nucleus
consists of one discrete state plus a continuum; this condition is
satisfied for the reaction with which we compare, \( ^{3}\mbox
{He}(e,e'p) \). Multiple discrete states would be straightforward to
implement. The spectral function used for \( ^{3} \)He comes from the
INFN/Rome group \cite{kiev97,salm97}.

A complete simulation consists of the following individual simulation runs: 

\begin{enumerate}
\item \emph{two-body breakup with external bremsstrahlung}. This run
  handles computation of the ``unradiated'' part of the cross section
  (see Sec.~\ref{sec:schw-corr}).  No internal bremsstrahlung is
  computed; rather, the computed PWIA cross sections are reduced by
  the Schwinger correction to account for the fraction which will be
  redistributed into the internal-bremsstrahlung tail. Sampling is
  performed in \( ({\bf p}_{e'},\theta _{p},\phi _{p}) \); the
  constraint of a definite \( (A-1) \) final-state mass provides the
  solution for \( p_{p'} \). External bremsstrahlung is allowed before
  the \( (e,e'p) \) vertex (modifying the beam energy) and afterwards
  (modifying the scattered-electron energy). The
  external-bremsstrahlung distribution is directly sampled, obviating
  the need for a cutoff correction.
\item \emph{two-body breakup with internal and external
    bremsstrahlung}. This run handles the part of the cross section
  which has been redistributed into the internal radiation tail. For
  each event, six variables are sampled \( ({\bf p}_{e'},{\bf p}_{p'})
  \).  First, external bremsstrahlung is computed along the incident
  electron direction (possibly modifying the incident electron
  energy). Then solutions are found for the radiated-photon energies
  corresponding to internal radiation along either the incident or
  scattered electron directions. The cross section is computed
  according to Eq.~(\ref{eq:bd-2b}). Finally, external bremsstrahlung
  is computed along the scattered electron direction (possibly
  modifying the detected electron energy). Here only the
  virtual-photon part \( (1-\delta _{v}) \) of the Schwinger
  correction is applied since the radiative tail corresponding to \(
  \delta _{r} \) is what we are computing.
\item \emph{continuum breakup with external bremsstrahlung}. This
  piece is similar to case 1 above, except that events are sampled in
  six kinematic variables \( ({\bf p}_{e'},{\bf p}_{p'}) \) since
  there is a continuum of possible \( (A-1) \) final states.
\item \emph{continuum breakup with both internal and external
    bremsstrahlung}. This simulation is similar to that of case 2
  above, except that due to the complete kinematic freedom in the
  final state, the photon energy is constrained only to be larger than
  \( \Delta E \) (the experimental resolution). Therefore,
  \emph{seven} variables are sampled, the radiated-photon energy being
  the seventh.
\item \emph{detection volume simulation}. This simulation is standard
  procedure for determining what fraction of the six-dimensional
  acceptance in \( ({\bf p}_{e'},{\bf p}_{p'}) \) can contribute to
  any given bin in a cross-section spectrum.
  The results of this piece are used to properly normalize
  the simulated spectra when producing cross section results.
  No energy losses effects are included, since this part of the
  simulation only measures the relative probability of detection of
  various kinematical configurations (regardless of their origin).
\end{enumerate}
The simulations, when properly weighted by sampling volumes and
numbers of trials, are combined to form simulated cross sections. The
cross sections can be plotted as the same sort of spectra shown in
experimental papers, by sorting the simulated events into histograms
in the same way an experimenter would sort data.

\section{Experimental Data}

The data with which we compare our simulations was acquired with the
three-spectrometer detector setup at the MAMI accelerator facility in
Mainz \cite{boeg95,blom98}.  Two of these spectrometers were used to
detect scattered electrons and knocked-out protons; the third served
as a luminosity monitor. A cryogenic gas target provided the \(
^{3}{\mathrm{H}e} \) target nuclei. The experiment measured cross
sections for the reaction \( ^{3}{\mathrm{H}e}(e,e'p) \) in a variety
of kinematic settings.  For more information on the experiment and its
physics goals, the reader can consult \cite{flor99}; here we focus
only on the essentials needed for the radiation-tail comparison. The
kinematical settings and experimental acceptances for the data
discussed here are given in Table \ref{tab:kinset}.
\begin{table}
{\centering \begin{tabular}{lr}
 Beam Energy &
 855 MeV \\
 Electron Scattering Angle &
 52.4 deg \\
 Scattered Electron Momentum (central) &
 627 MeV/\emph{c}\\
Momentum Acceptance&
\( \pm 9.5 \)\%\\
Nominal Electron Solid Angle&
20 msr\\
 Proton Detection Angle &
 -46.41 deg \\
 Proton Momentum (central) &
 661 MeV/\emph{c}\\
Momentum Acceptance&
\( \pm 7.4 \)\%\\
Nominal Proton Solid Angle&
4.8 msr\\
Experimental \( E_{m} \) Resolution (\( \Delta E \))&
0.4 MeV\\
\end{tabular}\par}

\caption{Kinematic Settings for Experimental Data}

{\par\centering \label{tab:kinset}\par}
\end{table}

One question which must be addressed in this study is ``how well can
the PWIA model describe the reaction?'' If the description is not
favorable, further work is useless since our model computes the basic
\( (e,e'p) \) cross section in PWIA.
\begin{figure}
  {\par\centering
    \resizebox*{0.4\textwidth}{!}{\includegraphics{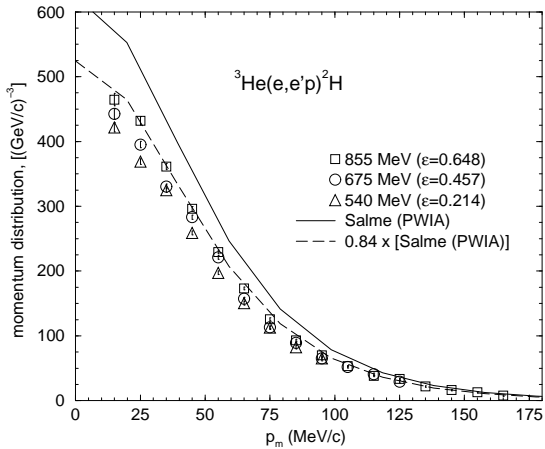}}
    \par}

\caption{Experimental and Theoretical Momentum Distributions for
  \protect\( ^{3}{\mathrm{H}e}(e,e'p)^{2}{\mathrm{H}}\protect \).
  Experimental distributions are shown for three different electron
  beam energies; the figure of 0.84 in the text refers to the ratio of
  the 855 MeV experimental distribution to the theoretical
  distribution.}

{\par\centering \label{fig:expmomdist}\par}
\end{figure}

Figure \ref{fig:expmomdist} shows the measured experimental momentum
distribution \( n_{\mathrm{exp}}=\sigma /(K\sigma _{ep}) \) where \( K
\) stands for the kinematical factors in Eq.~(\ref{eq:pw2b}). 
The experimental cross section $\sigma$ above has been radiatively
corrected using the traditional technique, which has been shown to
work well in this experiment \cite{flor99} for excitation energies
less than 20 MeV.
After radiative correction, the two-body peak could be cleanly
resolved from the continuum and a cross section assignment is
straightforward.

The momentum distribution is
compared to the theoretical two-body breakup spectral function
\cite{kiev97,salm97} \( S(S_{p},p_{m}) \).  \( S_{p} \) is the
single-proton separation energy and corresponds to \(
^{3}{\mathrm{H}e}\longrightarrow p\, \, +\, \, ^{2}{\mathrm{H}} \).
Aside from an overall scaling factor of 0.84, the theoretical spectral
function is in good agreement with the data. We interpret this
agreement as an attenuation of the outgoing proton flux in the
reaction, due to FSI, of constant magnitude 0.84; aside from this,
effects outside the PWIA are not important. Ref.~\cite{flor99} shows
several other instances of how, apart from this overall reduction,
PWIA calculations describe the data well. This good agreement can be
attributed to our use of a light nucleus (reducing FSI effects) and
the fact that our kinematics are directly tuned to the quasi-elastic
point, where PWIA should work best.

\section{Results}

We will discuss the results in two stages. First we will present
results using the unmodified Borie-Drechsel tail computation (\(
f_{\mathrm{mp}}=1 \)), including our extension for discrete states of
the \( (A-1) \) residual nucleus. These results show a clear
discrepancy in the tail region. We then present the derivation of our
tail correction factor \( f_{\mathrm{mp}} \). Then we present results
including this correction, which will be shown to resolve the
discrepancy.

Since the spectral function falls rapidly with \( p_{m} \), we were
concerned about relying on the theoretical momentum distribution
over the large $p_m$ acceptance of this experiment.
Eventual
discrepancies between our calculation and the experiment might
be due to inaccuracies in the hadronic structure of $^3\mathrm{He}$.
In order to reduce this possibility, we carried out this study
within a limited regime of $p_m$ by placing a cut on both the
experimental data and on the simulation results.
For all plots shown below,
only missing momenta in the range \( 40\leq p_{m}<50 \) MeV/\emph{c}
are considered (for both the experimental data and the simulation).
This particular region is near the top of the experimental acceptance
in \( p_{m} \), where we make the best measurement of the two-body
momentum distribution (on which the scaling factor is based)
and where we had the greatest statistical accuracy in the
experimental tail cross section.

\subsection{Results with Unmodified Tail Cross Section}

\begin{figure}
  {\par\centering
    \resizebox*{0.48\textwidth}{!}{\includegraphics{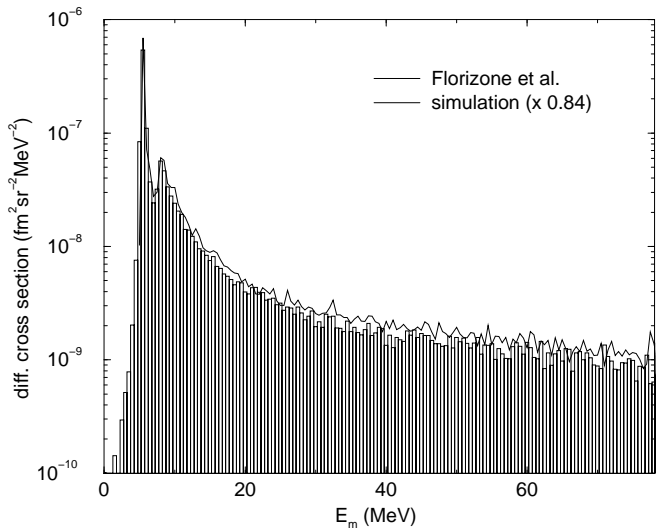}}
    \par}

\caption{Differential cross sections measured in the Mainz experiment 
  (histogram) and computed by the simulation program (solid line). The
  experimental data have not been corrected for bremsstrahlung
  effects. The multi-photon tail correction \protect\(
  f_{\mathrm{mp}}\protect \) has been set to
  unity.\label{fig:simvdata}}
\end{figure}

Fig.~\ref{fig:simvdata} shows a comparison of the measured cross
section, plotted as a function of the measured missing energy \( E_{m}
\), and the results of our simulation at the same kinematics. The
simulation result has been scaled by the factor 0.84 in accordance
with the findings for the momentum distribution.

\begin{figure}
  {\par\centering
    \resizebox*{0.48\textwidth}{!}{\includegraphics{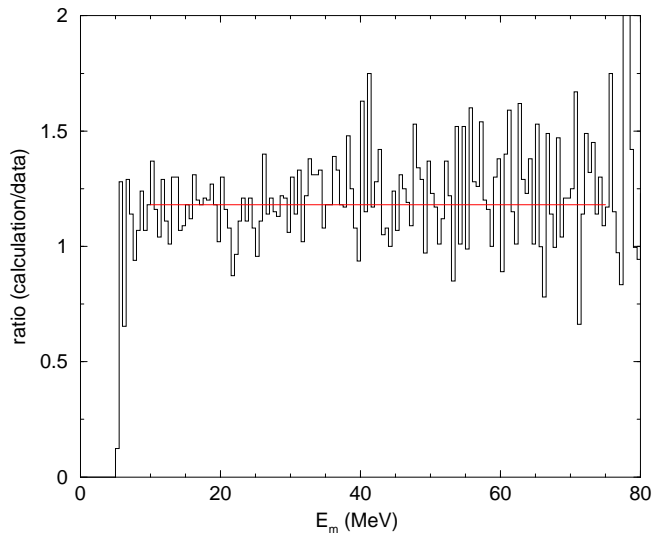}} \par}
\caption{\label{fig:simratio}Ratio of simulated 
  \protect\( (e,e'p)\protect \) spectrum (normalized by 0.84 as
  discussed in the text) to the experimentally measured cross section
  spectrum, plotted as a function of missing energy. A one-parameter
  fit to these data over the region \protect\(
  10<E_{m}<75\protect \) MeV yields a ratio of 1.18 (shown in
  the figure). The bin-to-bin fluctuations in the ratio are due to the
  statistical uncertainties in both the cross section and the
  simulation. The only significant deviation from the fit is below 10
  MeV, where the ratio decreases towards unity.}
\end{figure}

The agreement is generally excellent, the shape having been perfectly
reproduced within the statistical accuracy of the simulation. The
differences at the low-\( E_{m} \) side of the peak in the
spectrum are not really worrisome, since our simulation did not
include all possible mechanisms of energy loss and its accompanying
contribution to the experimental resolution. For example, while
external bremsstrahlung in the target-cylinder walls was accounted
for, ionization energy losses in this material (82 \( \mu \)m foil of
iron), for the incident and scattered electrons and ejected proton,
were not. Thus sharp features in the cross section (such as the low-\(
E_{m} \) peak) will not be correctly reproduced by the
simulation. At missing energies below about 10 MeV, where the
radiative tail is still a small contribution, the integrals
of the simulation and the data
agree to within 1\% (the integrals are not sensitive to
the shape differences discussed above).
This directly indicates that the empirical scaling factor is
applicable to the continuum breakup as well, since our scaling factor
of 0.84 was fixed by the two-body breakup results alone. For \(
E_{m} \) above 10 MeV, the simulation predicts a larger cross
section than observed, with an essentially constant excess of about
20\% (see Fig.~\ref{fig:simratio}). Since the shape reproduction is
excellent, and the strength in the low-\( E_{m} \) region is
well described, the comparison suggests a problem with the amplitude
of the calculated tail, but not with its shape.

\subsection{Radiation Tail Correction for Multiple-Photon Processes}

The Schwinger correction, which has been applied to the ``unradiated''
strength dominating the region of low \( E_{m} \), includes
the effects of multiple-photon emission and here the simulation and
data agree. The tail region cross section has been derived to first
order in real-photon emission, and here the simulation does not agree
with the measured data. Multiple photon processes are therefore
clearly indicated as a likely source of the discrepancy in the tail
strength.

A rigorous derivation of a multiple-photon tail cross section is
beyond the scope of this paper, but an intuitive derivation is easy to
provide. In the limit that the variation in the PWIA \emph{vertex}
cross section is very slow, bremsstrahlung processes only redistribute
strength with respect to the asymptotic kinematics. Thus if we add the
``unradiated'' part still residing in the peak to that residing in the
tail, we should recover the original PWIA cross section \( \sigma
_{\mathrm{pwia}} \). \( \delta _{r} \) represents the fraction of
strength radiated out of the peak, to first order. Thus if the
Borie-Drechsel cross section is valid in first order, its integral
will also yield a fraction \( \delta _{r} \) of \( \sigma
_{\mathrm{pwia}} \). However, the fraction remaining in the peak has
been adjusted to account for higher-order radiation; the cross section
here is \( e^{-\delta _{r}}\sigma _{\mathrm{pwia}} \). The sum of the
two is \( \left( e^{-\delta _{r}}+\delta _{r}\right) \sigma
_{\mathrm{pwia}} \).  The factor in parentheses is differs from unity
in second order. If we apply the multiplicative factor
\[
f_{\mathrm{mp}}=\frac{\left( 1-e^{-\delta _{r}}\right) }{\delta _{r}}\]
to the tail cross section, we recover \( \sigma _{\mathrm{pwia}} \) for the
sum of peak and tail cross sections in the presence of bremsstrahlung.

Note that this discussion only concerns the real-photon part
of the internal bremsstrahlung correction.
The external bremsstrahlung distribution described above is
an energy-loss distribution which includes the higher-order
contributions, thus they do not need to be considered here.

\subsection{Simulation including multi-photon tail correction}

\begin{figure}
  {\par\centering
    \resizebox*{0.48\textwidth}{!}{\includegraphics{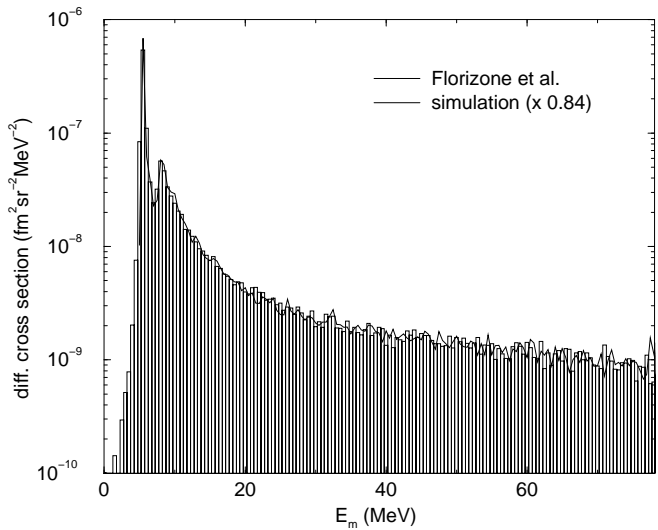}}
    \par}

\caption{Differential cross sections measured in the Mainz experiment 
  (histogram) and computed by the simulation program (solid line). The
  experimental data have not been corrected for bremsstrahlung
  effects. The full multi-photon tail correction \protect\(
  f_{\mathrm{mp}}\protect \) has been used.\label{fig:simvdata_corr}}
\end{figure}

\begin{figure}
  {\par\centering
    \resizebox*{0.48\textwidth}{!}{\includegraphics{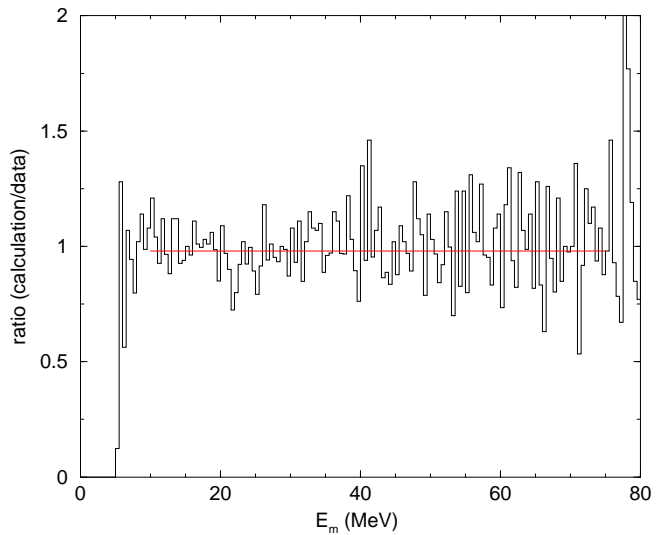}}
    \par}

\caption{Ratio of simulated \protect\( (e,e'p)\protect \) spectrum 
  (including the full tail correction and normalized by 0.84 as
  discussed in the text) to the experimentally measured cross section
  spectrum, plotted as a function of missing energy. A one-parameter
  fit to these data over the region \protect\(
  10<E_{m}<75\protect \) MeV yields a ratio of 0.98 (shown in
  the figure). The bin-to-bin fluctuations in the ratio are due to the
  statistical uncertainties in both the cross section and the
  simulation.\label{fig:ratio_corr}}
\end{figure}

The simulation was repeated including the multi-photon tail
correction, but otherwise identical (including the scaling factor
0.84). The results are presented in Figs.~\ref{fig:simvdata_corr} and
\ref{fig:ratio_corr}.

The reproduction of the shape of the tail is still excellent, which is
not surprising.  For the chosen kinematics, \( \delta _{r} \) has an
average value of 0.46, with a \( 1\sigma \) deviation of only 0.8\%
across the physical acceptance --- \( f_{\mathrm{mp}} \) is
essentially a multiplicative constant for the entire tail. However,
the simulation now reproduces the strength of the tail to the same
level of accuracy as for the peak region. These excellent results
provide unambiguous proof that multiple-photon processes are important
in the radiation tail, and also that our proposed correction factor is
valid at the few percent level (at least in the kinematical regime
studied here).

\begin{figure}
  {\par\centering
    \resizebox*{0.48\textwidth}{!}{\includegraphics{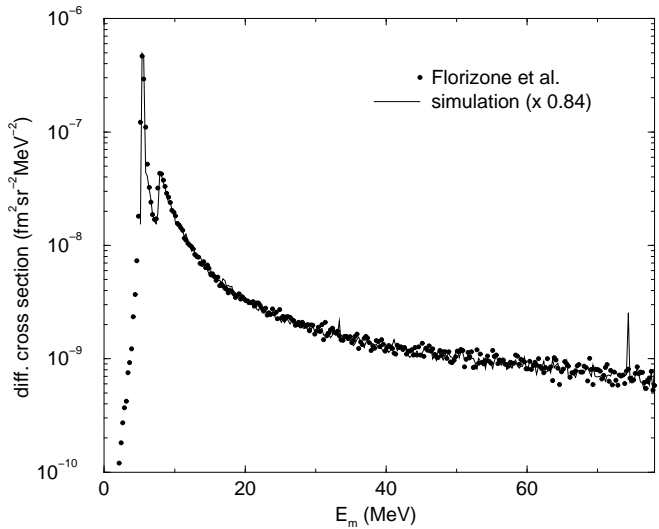}}
    \par}

\caption{\label{fig:simvdata_corr_full}Comparison of experimental data 
  to simulation (including tail correction). The plot shown is
  identical to that of Fig.~\ref{fig:simvdata_corr} except for the
  current plot, events (both experimental and simulated) with missing
  momenta \protect\( 30\leq p_{m}\leq 100\protect \) MeV/\emph{c} are
  included.  The bin size has been reduced to 0.25 MeV due to the much
  better statistical precision of these data.}
\end{figure}

Finally, we show in Fig.~\ref{fig:simvdata_corr_full} a similar
comparison of experimental and simulated cross-section spectra, except
here we consider an expanded range of \( p_{m} \). This check was
made to ensure that our agreement had not been fine-tuned for only
the small region of $p_m$ we had been considering.
The corresponding
ratio plot is very similar to Fig.~\ref{fig:ratio_corr}, with the
one-parameter fit yielding a ratio 0.976.  The same simulation
produced both Figs.~\ref{fig:simvdata_corr} and
\ref{fig:simvdata_corr_full}; the only difference was a change in the
\( p_{m} \) condition specified in the histogram-sorting program.

\subsection{Decomposition of Cross Section}

\begin{figure}
  {\par\centering
    \resizebox*{0.48\textwidth}{!}{\includegraphics{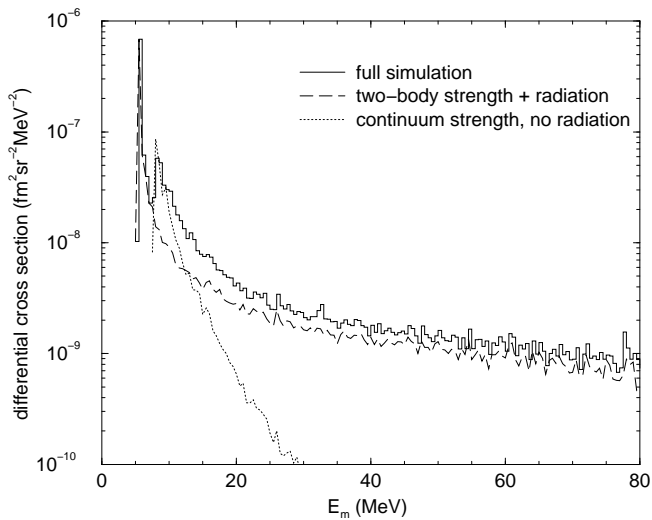}} \par}

\caption{\label{fig:pieces}Decomposition of simulated 
  \protect\( (e,e'p)\protect \) cross section. The solid curve gives
  the final result for the cross section including the entire spectral
  function and all radiative processes. The long-dashed curve only
  includes the two-body \protect\( ^{3}\mbox {He}(e,e'p)d\protect \)
  part of the spectral function, along with both classes of
  bremsstrahlung. The dotted curve includes only the continuum
  \protect\( ^{3}\mbox {He}(e,e'p)np\protect \) part of the spectral
  function, and the bremsstrahlung effects are not included.  When
  measuring in the continuum, the dotted curve is what one attempts to
  extract, and the long-dashed curve is physical background which must
  be removed by a radiative-correction procedure.}
\end{figure}

It is instructive to separate this cross section calculation into its
components.  This is a luxury that nature does not afford the
experimenter. One such decomposition is shown in
Fig.~\ref{fig:pieces}. Recall that what one usually wants to measure
is the cross section with the radiation tail removed. This corresponds
to the dotted line in Fig.~\ref{fig:pieces}, which is the simulation
result for continuum breakup \( ^{3}\mbox {He}(e,e'p)np \) with
bremsstrahlung turned off. The dashed curve shows the simulation for
\( ^{3}\mbox {He}(e,e'p)d \) only, but including the full
bremsstrahlung tail. The solid curve is the total simulation result as
shown in Fig.~\ref{fig:simvdata_corr}.

It is immediately apparent from the figure that there is no hope of
making a significant measurement for missing energies much above 15
MeV --- statistical fluctuations associated with the tail subtraction
procedure will render such a measurement insignificant. Most of the
observed cross section for \( E_{m}>20 \) MeV is due to \(
^{3}\mbox {He}(e,e'p)d \) reactions residing in the radiation tail. A
new experiment, made in a kinematical regime that does not result in
the generation of such a strong radiation tail, will be required to
make a statistically-significant measurement of the cross section in
this region.

We expect simulations such as that described here will become a
standard tool in the planning of experiments at large missing
energies, since one would clearly like to avoid performing an
experiment in kinematical regimes in which the radiation tail is
stronger than the cross section to be measured.

\section{Implications for Radiative Corrections of \lowercase{$(e,e'p)$} Data}

\begin{figure}
  {\par\centering
    \resizebox*{0.48\textwidth}{!}{\includegraphics{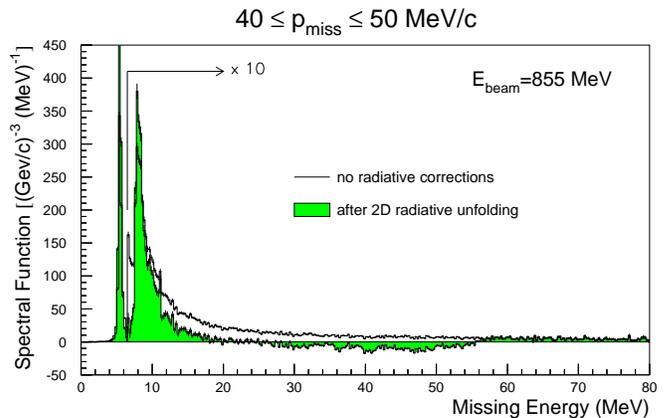}}

    \par}

\caption{\label{fig:badcor}Experimental data for 
  \protect\( ^{3}\mbox {He}(e,e'p)\protect \) before and after
  application of radiative corrections. In this figure, the
  experimental spectral function is displayed, which is related to the
  experimental cross sections by Eq.~(\ref{eq:sigpw}).}
\end{figure}

We stated in the introduction that this work was begun in an effort to
understand problems encountered when applying radiative corrections to
the \( ^{3}\mbox {He}(e,e'p) \) data discussed here.
Fig.~\ref{fig:badcor} (from Ref.~\cite{flor99}) illustrates the
problem.

The figure compares the measured spectral function (from the same
dataset which produced Fig.~\ref{fig:simvdata}) with the corresponding
spectral function after applying radiative corrections, \( i.e. \)
removing the radiation tail.  The tail correction procedure is
essentially identical to that of \cite{quin88} which was briefly
described in Sec.~\ref{sec:kineff}. For \( E_{m}>25 \) MeV, the
corrected spectral function is negative, clearly indicating a
deficiency.

In this case, the defect can not be obviously traced to
multiple-photon emission, since the tail computation is based on the
distribution of the exponential form of the Schwinger correction in \(
E_{m} \). Specifically, the number of expected experimental counts
inside a certain missing energy bin \(
E^{(0)}_{m}<E_{m}<E^{(0)}_{m}+\Delta E_{m} \) is given by
\[
  N_{\mathrm{exp}}(E^{(0)}_{m},\Delta E_{m})=
  e^{-\delta _{r}(\Delta E_{m})}N_{0}(E^{(0)}_{m})\, \, .
\]
  
Here \( N_{0} \) is the number of counts which would have been
measured in the absence of bremsstrahlung, and \( E^{(0)}_{m} \) is
the missing energy at which the counts would have appeared in the
absence of bremsstrahlung (or other processes which modify the
asymptotic particle energies such as ionization energy loss). As \(
\Delta E_{m} \) becomes larger, the bin includes more of the radiation
tail (the unradiated strength is included by definition) so that \(
N_{\mathrm{exp}} \) approaches \( N_{0} \). The distribution of the
radiation tail in \( E_{m} \) is thus
\[
  \frac{\partial N_{\mathrm{exp}}}{\partial E_{m}}=N_{0}\frac{\partial
  (e^{-\delta _{r}})}{\partial (\Delta E_{m})}\, \, .
\]
The factor \( e^{-\delta _{r}} \) again explicitly includes the
multiple-photon processes.

There are, however, several possible other reasons for the failure of
the radiation-correction procedure. We discuss them in detail in the
following sections.

\subsection{Incomplete Kinematic Reconstruction}

The radiative-correction process is usually applied in two kinematic
dimensions (see Ref.~\cite{flor99} for a thorough discussion). A
common choice for the two dimensions is \( (E_{m},p_{m}) \). First, a
two dimensional cross-section histogram is created with the
independent variables being \( E_{m} \) and \( p_{m} \).
Fig.~\ref{fig:adam1} would be appropriate if the \( z \)-axis
corresponded to cross section. As discussed in Sec.~\ref{sec:kineff},
the correction would begin at the left-hand edge of
Fig.~\ref{fig:adam1}. For each bin in \( p_{m} \) at this bin in \(
E_{m} \), correction factors would be applied and tails would be
generated and subtracted for the bins to the right. In such a
procedure, there is no information about any of the other kinematic
parameters; a swath in seven-dimensional kinematic space has been
reduced to a two-dimensional pixel.  The most common remedy for this
lack of information is to treat the entire bin as if the rest of the
parameters were fixed at their central values. However,
Fig.~\ref{fig:adam1} shows a substantial dependence in the tail
trajectories on the relative angle between \( {\bf q} \) and \( {\bf
  p}_{m} \). While both of these may vary across the detector
acceptances, both are held fixed in the correction procedure. \( |{\bf
  q}| \) is also held fixed, while we know that it varies
substantially across the acceptances and causes large changes in the
cross section through \( \sigma _{ep} \). For example, at the
kinematics corresponding to Fig.~\ref{fig:badcor}, \( \delta
(Q^{2})/\mu (Q^{2}) \) (the standard deviation of \( Q^{2} \) divided
by the mean value) is about 7.6\% for the pixel \( 5.5<E_{m}<5.9 \)
MeV, \( 40<p_{m}<50 \) MeV/\emph{c.} This leads to a 15\% RMS
variation of the cross section due to \( \sigma _{ep} \) \emph{alone.}
The variation in the computed \( ^{3}\mbox {He}(e,e'p) \) cross
section is about 20\%.

Procedures have been developed to perform the correction in more
dimensions (\emph{e.g.}~four were used in Ref.~\cite{flor99}) but such
schemes are only feasible for experiments with good statistical
precision, as each additional dimension tends to reduce the
statistical precision per bin by roughly an order of magnitude.

\subsection{Simplifying Assumptions About the Radiation Tail}

The standard correction procedure uses a derivative of the Schwinger
correction factor \emph{vs.} \( E_{m} \) to generate the tail
distribution. However, as explained in Secs.~\ref{sec:kineff} and
\ref{sec:comput}, there are two directions this tail can take in \(
(E_{m},p_{m}) \) space, corresponding to the two terms of the peaking
approximation. The Schwinger correction gives no guidance as to how
much strength resides in each tail. The standard practice is to assume
\cite{quin88,holt95}:

\begin{itemize}
\item the incoming and outgoing electrons contribute independently to
  the tail, so one may factor \( e^{-\delta }=C_{e}(\Delta E_{m})\,
  C_{e'}(\Delta E_{m}) \);
\item the two contributions are equal, so \( C_{e}=C_{e'}=e^{-\delta
    /2} \).
\end{itemize}
The Borie-Drechsel formula (Eq.~(\ref{eq:bd-cont})) for the radiation
tail clearly does not have these properties. Firstly, the two tails
add instead of multiply. Secondly, they are not equal. Even for
vanishingly small photon energies (\( k_{e} \) and \( k_{e'} \)), the
two terms differ by the factors \( \ln \left( 2E_{e}/m_{e}\right) \)
\emph{vs.} \( \ln \left( 2E_{e'}/m_{e}\right) \) (which have values
8.12 and 7.81 respectively in the kinematics studied here). As the
radiated-photon energy increases, the difference between the two terms
also increases; for a photon energy of 100 MeV, the incident-electron
contribution is 6\% larger than that of the scattered electron in the
present kinematics. The difference between the tail magnitudes is
mainly driven by the ratio $\omega/E_e$; when it is large, the
tail strengths differ more.
For a specific experiment planned at JLab with $\omega = 834$ MeV
and $E_e = 1245$ MeV, the two tails differ by about 16\% in strength.

\subsection{Comparison With Direct Tail Simulation}

The radiation tail calculation presented here suffers from none of the
above deficiencies.

\begin{itemize}
\item the tail is generated event-by-event, so for each tail
  evaluation, the complete kinematic information is available
\item the distribution of the tail strength in this kinematic space is
  based on a first-order QED calculation, not on plausible
  assumptions.
\end{itemize}
The main deficiency of our computation is the nature of the
multi-photon correction factor. It is based on arguments of
probability conservation rather than on a rigorous QED calculation.
This argument is however of the same type which leads to the
exponentiation of \( \delta _{r} \) in the standard approach.  A
critical review of including higher-order terms via exponentiation
(including a summary of relevant literature) can be found in
\cite{maxi69}.

\subsection{An Improved Radiative Correction Procedure}

The findings reported above indicate that the standard radiative
correction procedure is not likely to work for cases in which the
detector acceptances are relatively large (producing large variations
in \( Q^{2} \) for individual pixels in cross-section histograms) or
in which a correction is being made over a large range in \( E_{m} \)
(so that the differences between the two tails becomes important). Our
findings suggest an improved method for radiatively ``correcting''
experimental data.

The procedure would begin with a model spectral function and an
accurate model of the experimental apparatus, such as has been
described here. A simulation code similar to ours should be used to
generate a ``radiated'' cross section spectrum. A comparison between
the experimental and simulated histograms will indicate regions of
discrepancy, and the discrepancy function can be used to modify the
model spectral function. The procedure is repeated until it converges,
at which point the model spectral function corresponds to the
unradiated result.  This procedure is independent of the PWIA if we
replace the theoretical spectral function described above with the
``distorted'' spectral function \cite{kell96}.  We learned during the
final stages of preparing this article that such an iterative
procedure has been developed and successfully applied for an
experiment in Hall C at Jefferson Lab \cite{dutt99}.

\section{Further Work}

It is desirable to have a more rigorous theory provide the
multi-photon tail cross section. The beginnings of such an approach
can be found in \cite{maki94}.  We encourage this group to complete
and publish these results, especially since they have made some
detailed evaluations of their approach in the Jefferson Lab energy
domain.

It would also be interesting to reanalyze some of the older high-\(
E_{m} \) \( (e,e'p) \) data using the improved technique described
here, since these data have been a source of controversy, given the
sometimes puzzling behavior of the cross section at high missing
energies. The current study indicates that some of this cross section
might well be misidentified bremsstrahlung strength.

\section{Conclusions}

We have presented a framework for computing \( (e,e'p) \) cross
sections which includes the radiation tail to first order. The
computed cross sections have been compared to experimental data in
such a way that effects such as acceptance averaging are correctly
accounted for, allowing a direct evaluation of the radiation tail
cross section calculation.

The computed tail reproduces both the shape and magnitude of the
experimental spectrum perfectly within experimental errors. It was
necessary to derive a correction, applied to the radiation tail, for
higher-order bremsstrahlung effects before this agreement could be
obtained; the original tail calculation treated bremsstrahlung only to
first order.

A standard radiative correction procedure has also been applied to
these data.  Such a procedure is designed to move the radiation tail
strength back into the originating kinematic bins. The straightforward
application of this procedure (that is, without tweaking parameters to
improve the agreement) results in physically unreasonable
``deradiated'' cross sections in the tail-dominated part of the
spectrum. There are several reasons to expect such a failure. An
obvious one is that this procedure collapses a complicated kinematical
hypersurface (along which the cross section varies substantially) to a
single point in \( (E_{m,}p_{m}) \).  More subtle are the disturbing
differences between the properties of the correction-procedure tails
and those of a tail cross section rigorously computed in QED. The
observed flaws in the correction procedure are not likely to affect
earlier data taken at low missing energies, \emph{e.g.,} at NIKHEF,
Bates, and Mainz. They may affect earlier high-\( E_{m} \) data, and
will likely be fatal for several of the \( (e,e'p) \) experiments
planning to measure at large-\( E_{m} \) at Jefferson Lab. Our results
indicate how radiative corrections should be applied so as to avoid
such problems.

The current project has yielded quantitative illustrations of the
failure of the standard radiative-correction procedure for \( (e,e'p)
\) experiments.  We have also shown that a simulation, coupled with an
accurate model for the radiative-tail cross section, can radiate the
theory (instead of deradiating the data) and achieve excellent
agreement with experiment in a situation where the correction
procedure fails. The simulation technique described here provides a
basis for iterative radiation-correction procedures for future \(
(e,e'p) \) experiments.

\appendix

\section{Other Limitations of the PWIA}

In Sec.~\ref{sec:eephen} we were careful to distinguish the
spectral-function quantities \( \epsilon \) and \( \bbox {\rho } \)
from the experimentally determined values \( E_{m} \) and \( {\bf
  p}_{m} \). Even in the absence of bremsstrahlung this is necessary
since the PWIA never holds completely, and is sometimes grossly
violated. For such cases, \( \epsilon \neq E_{m} \) and \( \bbox {\rho
  }\neq -{\bf p}_{m} \). Final State Interactions (FSI) between the
ejected proton and the residual nucleus provide an illustrative
example of how the correspondence is broken. At the photon-proton
vertex of Fig.~\ref{fig:feyn1}, the amplitude for the interaction will
depend on the particular values of \( \epsilon \) and \( \bbox {\rho }
\). A subsequent interaction between the ejected proton and the
residual nucleus can change both the momentum and excitation energy of
the residual system, thus leading (through Eq.~(\ref{eq:fv_r})) to
values for \( E_{m} \) and \( {\bf p}_{m} \) different than \(
\epsilon \) and \( \bbox {\rho } \). While this point is not
particularly relevant for the present work, we mention it here for
completeness and because it is apparently often overlooked.

\section*{Acknowledgements}

This work was supported in part (JAT) by a grant from the U.S.
National Science Foundation.

\end{document}